\documentclass{article}
\usepackage{spconf,amsmath,graphicx,verbatim,multirow}

\title{Non-native Speaker Verification for Spoken Language Assessment}
\name{Linlin Wang$^\dagger$, Yu Wang$^\dagger$, Mark J. F. Gales\thanks{This paper reports on research supported by Cambridge Assessment, University of Cambridge. Thanks to Cambridge English Language Assessment for support and access to the BULATS and Linguaskill data. $^\dagger$ Both authors contributed equally. The authors would also like to thank Dr Kate Knill and Dr Anton Ragni for valuable discussions during the preparation of this manuscript.}}
\address{ALTA Institute / Engineering Department, Cambridge University, UK}

\begin{document}
%
\maketitle
\begin{abstract}
Automatic spoken language assessment systems are becoming more popular in order to handle increasing interests in second language learning. One challenge for these systems is to detect malpractice. Malpractice can take a range of forms, this paper focuses on detecting when a candidate attempts to impersonate another in a speaking test. This form of malpractice is closely related to speaker verification, but applied in the specific domain of spoken language assessment. Advanced speaker verification systems, which leverage deep-learning approaches to extract speaker representations, have been successfully applied to a range of native speaker verification tasks. These systems are explored for non-native spoken English data in this paper. The data used for speaker enrolment and verification is mainly taken from the BULATS test, which assesses English language skills for business. Performance of systems trained on relatively limited amounts of BULATS data, and standard large speaker verification corpora, is compared. Experimental results on large-scale test sets with millions of trials show that the best performance is achieved by adapting the imported model to non-native data. Breakdown of impostor trials across different first languages (L1s) and grades is analysed, which shows that inter-L1 impostors are more challenging for speaker verification systems.
\end{abstract}
\begin{keywords}
speaker verification, non-native speech
\end{keywords}
\section{Introduction}
\label{sec:intro}

\begin{figure}[ht!]
  \centering
  \includegraphics[scale=0.18]{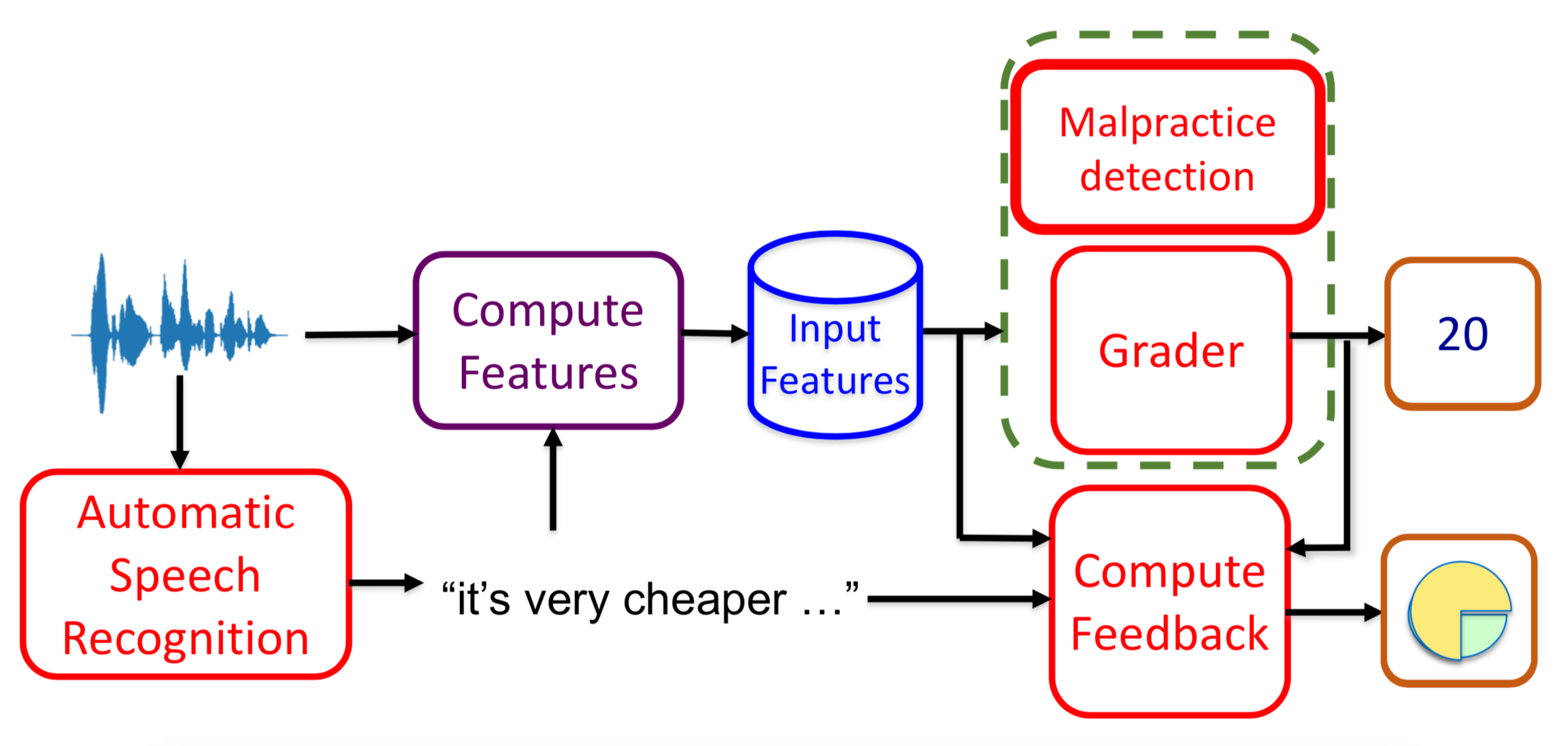}
  \caption{Diagram of an automatic spoken language assessment system.}
  \label{fig:diagram}
\end{figure}

Automatic spoken assessment systems are becoming increasingly popular, especially for English with the high demand around the world for learning of English as a second language \cite{Zechner2009,Witt2000,Metallinou2014,Wang2018a}. In addition to assessing a candidate's English ability such as fluency and pronunciation and giving feedback to the candidate, these automatic systems also need to ensure the integrity of the candidate's score by detecting malpractice, as shown in Figure~\ref{fig:diagram}. Malpractice is the action by a candidate that breaks the assessment regulation and potentially threatens the reliability of the exam and associated certification. Malpractice can take a range of forms in spoken language assessment scenarios, such as using or trying to use unauthorised materials, impersonation, speaking irrelevant to prompts/questions, speaking in his/her first language (L1) instead of the target language for spoken tests, \textit{etc}. This work aims to investigate the problem of automatically detecting impersonation, in which a candidate attempts to impersonate another in a speaking test. This is closely related to speaker verification.

\vspace{-2ex}

Speaker verification is the process to accept or reject an identity claim by comparing the speaker-specific information extracted from the verification speech with that from the enrolment speech of the claimed identity. These approaches can be directly applied to detect impersonation in spoken language tests. The performance of speaker verification systems has advanced considerably in the last decade with the development of i-vector modelling \cite{Dehak2011}, in which a speech segment or a speaker is represented as a low-dimensional feature vector. Extraction of i-vectors is normally based on a Gaussian mixture model (GMM) based universal background model (UBM). This fixed length representation can then be used with a probabilistic linear discriminant analysis (PLDA) model to produce verification scores by comparing speaker representations, which are then used to make valid or impostor speaker decisions \cite{Prince2007,Kenny2010,Garcia2011,Garcia2012}. Recently, with developments in deep learning, performance of speaker verification systems has been improved by replacing the GMM with a deep neural network (DNN) to derive statistics for extracting speaker representations. This DNN is usually trained to take a fixed length window of the acoustics and discriminate between speakers using supplied speaker labels as targets. To handle the variable-length nature of the acoustic signal, a pooling layer is used to yield the final fixed-dimensional speaker representation. In \cite{Variani2014}, a DNN was trained at the frame level, and pooling was performed by averaging activation vectors of the last hidden layer over all frames of an input utterance. In \cite{Snyder2016,Snyder2017,Snyder2018}, segment-level embeddings were extracted, which are referred to as x-vectors \cite{Snyder2018} with data augmentation. By leveraging data augmentation based on background noise and acoustic reverberation, these x-vectors based systems can achieve better performance than i-vector and d-vector based systems on standard speaker verification tasks.

There has been some previous work on tasks related to non-native speech data using speaker verification approaches, such as detection of non-native speech \cite{Shriberg2008}, classification of native/non-native English \cite{Tan2010} and L1 detection \cite{Omar2010}. In \cite{Qian2016}, meta-data (L1) sensitive bottleneck features were employed within the i-vector framework to improve the performance of speaker verification with non-native speech. In contrast, this paper focuses on making use of the state-of-the-art deep-learning based speaker verification approaches to detect candidate impersonation in an English speaking test. As there is limited amounts of data available for the non-native learner task, it is of interest to investigate adapting a standard speaker verification task to this non-native task. Here a system based on the VoxCeleb dataset \cite{Nagrani2017,Chung2018} is adapted to the BULATS task. Two forms of adaptation are examined: modifying the PLDA distance measure; and adapting the process for extracting the speaker representation by ``fine-tuning"  the network to the target domain.  Furthermore, detailed analysis of performance is also done with respect to speaker attributes. Gender is an important attribute in impostor selection for standard speaker verification tasks, and for non-native speech, there are two additional speaker attributes: the L1 and the language proficiency level\footnote{Language ability level is referred to as ``grade" in this work.}, which should also be taken into consideration for speaker verification. 

This paper is organised as follows. Section 2 gives an overview of speaker verification systems, and Section 3 introduces the non-native spoken English corpora used in this work. Experimental setup is described in Section 4, results and analysis are detailed in Section 5, and finally, conclusions are drawn in Section 6.

\section{Speaker Verification Systems}

In this work both i-vector and x-vector representations are used. For the i-vector speaker representation the form described in  \cite{Dehak2011,Povey2011} is used. This section will just discuss the x-vector speaker representation as this is the form that is adapted to the non-native verification task. 



\subsection{Deep neural network embedding extractor}\label{sec:x-vector model}
There are three blocks to form the DNN for extracting the utterance-level speaker representation, or embedding. The first block of the deep embedding extractor is a frame-level feature extractor. The input to this block is a sequence of acoustic feature vectors $\{\mathbf{x}_{1},\mathbf{x}_{2},\cdots\mathbf{x}_{T}\}$ of $T$ frames. This part normally consists of a number of hidden layers such as long short-term memory (LSTM) \cite{Heigold2016} or time delay neural network (TDNN) layers \cite{Snyder2017,Snyder2018}. The activations of the last hidden layer of this block for the input frames, $\{\mathbf{h}_{1},\mathbf{h}_{2},\cdots\mathbf{h}_{T}\}$, form the input to the second block which is a statistics pooling layer. This layer converts variable-length frame-level features into a fixed-dimensional vector by calculating the mean vector, $\boldsymbol{\mu}$ and standard deviation vector $\boldsymbol{\sigma}$ of the frame-level feature vectors over the $T$ frames.
The third block takes the statistics as the input and produces utterance-level representations using a number of stacked fully-connected hidden layers. The output of the DNN extractor is a softmax layer, and each of the nodes corresponds to one speaker identity. This DNN extractor is trained based on a cross-entropy loss function using the supplied speaker labels to get the targets. Consider there are $N$ training segments and $S$ speakers, the cross-entropy can be written as
\begin{equation}
\mathcal{F}\left(\boldsymbol{\theta}\right)=-\sum_{n=1}^N\sum_{k=1}^K\delta\left(s,s_{k}^{\left(n\right)}\right)\text{log}P\left(s|\mathbf{x}_{1:T}^{\left(n\right)},\boldsymbol{\theta}\right),
\end{equation}
where $\boldsymbol{\theta}$ represents the parameters of the DNN and  $\delta\left(\cdot\right)$ represents the Kronecker delta function. $s_{k}^{\left(n\right)}$ represents that the speaker label for segment $n$ is $s_{k}$. After the DNN is trained, the utterance-level embeddings, $\mathbf{e}_{d}$, are normally extracted from the output of the affine component that is with or without the nonlinear activation function applied of one hidden layer in the third block of the DNN \cite{Snyder2017,Snyder2018}.

\subsection{PLDA classifier and adaptation}
After the speaker embeddings are extracted, they are used to train a PLDA model that yields the score (distance) between speaker embeddings. The training of the PLDA models aims to maximise the between-speaker difference and minimise the within-speaker variation, typically using expectation maximisation (EM). A number of variants of PLDA models have been introduced into the speaker verification task based on this ``standard" PLDA \cite{Prince2007}: two-covariance PLDA \cite{Brummer2010} and heavy-tailed PLDA \cite{Kenny2010}. The variant implemented in the Kaldi toolkit \cite{Povey2011}, and used in this work, follows \cite{Ioffe2006} and is similar to the two-covariance model. This model can be written as 
\vspace{-0.1cm}
\begin{align}
\mathbf{e} & =\mathbf{\boldsymbol{y}}+\mathbf{z},\\
p\left(\mathbf{y}\right) & =\mathcal{N}\left(\mathbf{y};\mathbf{0},\boldsymbol{\Gamma}\right),\\
p\left(\mathbf{e}|\mathbf{y}\right) & =\mathcal{N}\left(\mathbf{e};\mathbf{\mathbf{y}}+\boldsymbol{\mu},\boldsymbol{\Lambda}\right),
\end{align}
where $\mathbf{e}$ is the speaker embedding. The vector $\mathbf{y}$ represents the underlying speaker vector and $\boldsymbol{\mu}$ represents its mean. $\mathbf{z}$ is the Gaussian noise vector. For speaker verification tasks, estimation of this PLDA model can be performed by estimating the between-speaker covariance matrix, $\boldsymbol{\Gamma}$, and within-speaker covariance matrix, $\boldsymbol{\Lambda}$, using the EM algorithm.
 
PLDA is a powerful approach to classifying speakers given a large amounts of training data with speaker labels \cite{Garcia2014,Garcia2014a,Villalba2014}. However, large amounts of labelled training data may not be available in the domain of interest such as the one considered in this paper, the non-native speaker verification. One approach to alleviate this problem is to do adaptation from a pre-trained out-of-domain model to the target domain. There are a number of methods for adapting the PLDA model in both supervised and unsupervised manners \cite{Garcia2014b,Villalba2014}. The Kaldi toolkit implements an unsupervised adaptation method which does not require knowledge of speaker labels  \cite{Povey2011}. This method aims at adapting $\boldsymbol{\Gamma}$ and $\boldsymbol{\Lambda}$ of the out-of-domain PLDA model to better match the total covariance of the in-domain adaptation data.


\section{Non-native Spoken English Corpora}
\label{sec:corpora}
The Business Language Testing Service (BULATS) test of Cambridge Assessment English~\cite{chambers-2011-bulats} is a multi-level computer-based English test. It consists of read speech and free-speaking components, with the candidate responding to prompts. The BULATS spoken test has five sections, all with materials appropriate to business scenarios. The first section (A) contains eight questions about the candidate and their work. The second section (B) is a read-aloud section in which the candidates are asked to read eight sentences. The last three sections (C, D and E) have longer utterances of spontaneous speech elicited by prompts. In section C the candidates are asked to talk for one minute about a prompted business related topic. In section D, the candidate has one minute to describe a business situation illustrated in graphs or charts, such as pie or bar charts. The prompt for section E asks the candidate to imagine they are in a specific conversation and to respond to questions they may be asked in that situation (e.g. advice about planning a conference). This section is made up of 5x 20 seconds responses. 

Each section is scored between 0 and 6; the overall score is therefore between 0 and 30. This score is then mapped into Common European Framework of Reference (CEFR)~\cite{CEFR2001} language proficiency levels, which is an international standard for describing language ability on a six-level scale. Each candidate is finally assigned a ``grade", ranging from minimal (A1) and basic (A2) command, through limited but effective (B1) and generally effective (B2) command, to good operational (C1) and fully operational (C2) command of the spoken language.

In this work, non-native speech from the BULATS test is used as both training and test data for the speaker verification systems. To investigate how the systems generalise, data for testing is also taken from the Cambridge Assessment English Linguaskill~\footnote{https://www.cambridgeenglish.org/exams-and-tests/linguaskill/} online test. Like BULATS, this is also a multi-level test and has a similar format composed of the same five sections as described before but assesses general English ability.



\section{Experimental Setup}
A set of 8,480 candidates from BULATS was used for training. The approximately 280 hours of speech covers a wide range of more than 70 different L1s. 
There are 15 major L1s with more than 100 candidates for each, including Tamil, Gujarati, Hindi, Telugu, Malayalam, Bengali, Spanish, Russian, Kannada, Portuguese, French, \textit{etc}. Data augmentation was applied to the training set, and each recording was processed with a randomly selected source from ``babble", ``music", ``noise" and ``reverb"~\cite{Snyder2018}, which roughly doubled the size of the original training set. 
Another set of 8,318 BULATS candidates was used as one test set to evaluate the system performance. There are 7 major L1s in this set, each of which has more than 100 candidates: Spanish, Thai, Tamil, Arabic, Vietnamese, Polish and Dutch. There are no overlapping candidates between the BULATS training and test sets. 
The other test set of 2,540 candidates came from the Linguaskill test, of which there are 6 major L1s each with more than 100 candidates: Hindi, Portuguese, Japanese, Spanish, Thai and Vietnamese.
Each of the training set and two test sets was fairly gender balanced, with approximately one third of candidates graded as B1, one third graded as B2, and the rest graded as A1, A2, C1, or C2, according to CEFR ability levels. 
For each test set candidate, responses from sections A and B were used for speaker enrolment (approximately 180s), while the more challenging free-speaking sections C, D, and E were used for whole section-level verification (approximately 60s for each section).

\section{Experimental results}
\label{sec:results}

\subsection{Baseline system performance}\label{sub:gender_dependent}

Gender is generally considered an important speaker attribute, and impostor trials were first selected from the same gender group as the reference speaker, as commonly done in standard speaker verification tasks. This resulted in a total of 104.8 million verification trials for the BULATS test set and 9.7 million trials for the Linguaskill test set.

An i-vector/PLDA system and an x-vector/PLDA system were first trained on the ``in-domain" BULATS training set. For the i-vector system, 13-dimensional perceptual linear predictive (PLP) features were extracted using the HTK toolkit~\cite{Young2015_htk} with a frame-length of 25ms. A UBM of 2,048 mixture components was first trained with full-covariance matrices, and then 600-dimensional i-vectors were extracted for both training and test sets. For the x-vector system, 40-dimensional filterbank features were also extracted using HTK with a frame-length of 25ms. DNN configurations were the same as used in~\cite{Snyder2018}, and 512-dimensional x-vectors were extracted from the affine component of the segment-level layer immediately following the statistics pooling layer.  

Performance of the two baseline systems is shown in Table~\ref{tab:bulats-based-systems} in terms of equal error rate (EER). The x-vector system yielded lower EERs on both BULATS and Linguaskill test sets. 

\begin{table}[ht!]
  \caption{\% EER performance of BULATS-trained baseline systems on BULATS and Linguaskill test sets.}
  \label{tab:bulats-based-systems}
  \centering
  \vspace{1.5ex}
  \begin{tabular}{ccc}
    \hline
    System      & BULATS   & Linguaskill \\
    \hline
	BULATS i-vector/PLDA & 0.69 & 0.72 \\
	BULATS x-vector/PLDA & 0.66 & 0.70 \\
    \hline
  \end{tabular}
\end{table}

In addition to the models trained on the BULATS data, it is also interesting to investigate the application of ``out-of-the-box" models for standard speaker verification tasks to this non-native speaker verification task as there is limited amounts of non-native learner English data that is publicly available. In this paper, the Kaldi-released~\cite{Povey2011} VoxCeleb x-vector/PLDA system was used as imported models, which was trained on augmented VoxCeleb 1~\cite{Nagrani2017} and VoxCeleb 2~\cite{Chung2018}. There are more than 7,000 speakers in the VoxCeleb dataset with more than 2,000 hours of audio data, making it the largest publicly available speaker recognition dataset. 30 dimensional mel-frequency cepstral coefficients (MFCCs) were used as input features and system configurations were the same as the BULATS x-vector/PLDA one. It can be seen from Table~\ref{tab:voxceleb-based-systems} that these out-of-domain models gave worse performance than baseline systems trained on a far smaller amount of BULATS data due to domain mismatch. Thus, two kinds of in-domain adaptation strategies were explored to make use of the BULATS training set: PLDA adaptation and x-vector extractor fine-tuning. For PLDA adaptation, x-vectors of the BULATS training set were first extracted using the VoxCeleb-trained x-vector extractor, and then employed to adapt the VoxCeleb-trained PLDA model with their mean and variance. For x-vector extractor fine-tuning, with all other layers of the VoxCeleb-trained model kept still, the output layer was re-initialised using the BULATS training set with the number of targets adjusted accordingly, and then all layers were fine-tuned on the BULATS training set. Here the PLDA adaptation system is referred to as \textbf{X1} and the extractor fine-tuning system is referred to as \textbf{X2}. Both adaptation approaches can yield good performance gains as can be seen from Table~\ref{tab:voxceleb-based-systems}. PLDA adaptation is a straightforward yet effective way, while the system with x-vector extractor fine-tuning gave slightly lower EERs on both BULATS and Linguaskill test sets by virtue of a relatively ``in-domain" extractor prior to the PLDA back-end.

\begin{table}[ht!]
  \caption{\% EER performance of VoxCeleb-based systems on BULATS and Linguaskill test sets.}
  \label{tab:voxceleb-based-systems}
  \centering
  \vspace{1.5ex}
  \begin{tabular}{lcc}
    \hline
    System     & BULATS   & Linguaskill \\
    \hline
VoxCeleb x-vector/PLDA & 0.85 & 1.44 \\
+ PLDA adaptation (X1) & 0.55 & 0.62 \\
+ Extractor fine-tuning (X2) & 0.49 & 0.55 \\
    \hline
  \end{tabular}
\end{table}

\begin{figure}[ht!]
  \centering
  \includegraphics[scale=0.52]{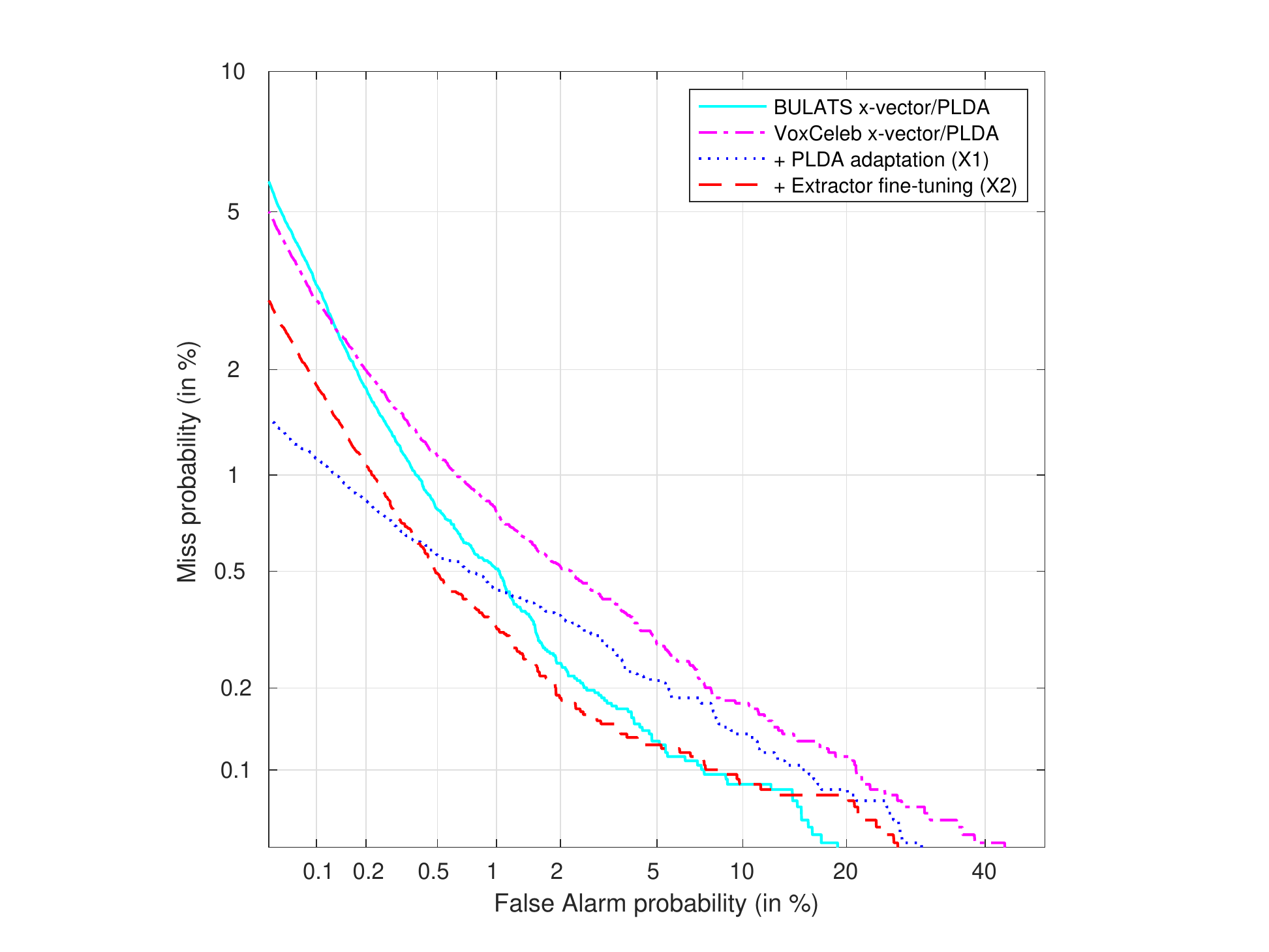}
  \caption{DET curves of the four x-vector/PLDA systems on the BULATS test set with impostors from the same gender group as the reference speaker.}
  \label{fig:gender_DET}
  \vspace{-0.2cm}
\end{figure}

Detection Error Tradeoff (DET) curves of the four x-vector/PLDA systems on the BULATS test set were illustrated in Figure~\ref{fig:gender_DET}. It can be seen that, both adaptation systems outperformed the original VoxCeleb-trained system in any threshold of the false alarm (FA) probability and the miss (MS) probability. The extractor fine-tuning system only gave higher MS probability than the PLDA adapted one with FA probability below 0.4\%, while for a large range of FA probabilities above 0.4\%, the extractor fine-tuning system outperformed the PLDA adapted one. 


Furthermore, by leveraging the large-scale VoxCeleb dataset, both adaptation systems produced lower EERs than baseline systems solely trained on BULATS data, especially the extractor fine-tuning one, which gave a reduction rate of 26$\%$ in EER over the baseline x-vector/PLDA system on the BULATS test set. It can also be seen from Figure~\ref{fig:gender_DET} that, the extractor fine-tuning system gave consistently better performance than the baseline systems for almost any threshold of FA and MS. 

\subsection{Impostor attributes analysis}
\label{sec:analysis}

As mentioned in Section~\ref{sub:gender_dependent}, gender is an important attribute when selecting impostors. For the non-native English speech data considered in this work, there are two additional attributes that may significantly impact performance, the candidate speaking ability (grade) and L1. In this section, the impact of both attributes on verification performance is analysed on the BULATS test set using the extractor fine-tuning system (X2) detailed in Section~\ref{sub:gender_dependent} with impostors selected from the same gender group as the reference speaker. Taking EER as the operating threshold, both grade and L1 breakdown are investigated with respect to the number of impostor trials resulting in false alarm (FA) errors. 

As there were only a small number of speakers graded as C1 or C2 in the BULATS test set, the two grade groups were merged into one group as C in the following analysis. Also for a fair comparison, 200 speakers were randomly selected (roughly gender balanced) for each grade group from the BULATS test set, and the grade breakdown is shown in Table~\ref{tab:grade_breakdown}. For lower grades, impostor trials from the grade group of A1 dominated FA errors as A1 speakers tend to speak short utterances, which is more challenging for the systems. For higher grades (B2 and C), impostor trials from the grade group of C constituted a larger portion of FA errors probably due to the fact that C speakers tend to speak long utterances in a more ``native" way and they are also similar to B2 speakers.  

\vspace{-2ex}
\begin{table}[ht!]
  \caption{Grade breakdown of the percentage of impostor trials with FA errors at the operating threshold of EER for the extractor fine-tuning system on a subset of the BULATS test set.}
  \label{tab:grade_breakdown}
  \centering
  \vspace{1.5ex}
  \begin{tabular}{cccccc}
    \hline
	Grade & \multicolumn{5}{c}{Grade of Impostor Spkr.} \\
	\cline{2-6}
	 Ref. Spkr.  & A1 & A2 & B1 & B2 & C \\
	 \hline
	 A1 & 65.8 & 27.5 & 5.8 & 0.3 & 0.6 \\
	 A2 & 60.9 & 29.9 & 7.1 & 0.9 & 1.3 \\
	 B1 & 46.5 & 26.8 & 13.1 & 7.6 & 5.9 \\
	 B2 & 11.4 & 11.9 & 19.2 & 25.9 & 31.7 \\
	 C  & 17.7 & 12.0 & 10.3 & 24.3 & 35.6 \\
    \hline
  \end{tabular}
\end{table}

The numbers of speakers from different L1 groups also varied in the BULATS test set. For a fair comparison, 200 speakers were randomly selected (roughly gender balanced) for each of 6 major L1s. The L1 breakdown is shown in Table~\ref{tab:L1_breakdown}, where impostor trials from the same L1 group as the reference speaker generally dominated FA errors. English learners from the same L1 group tend to have similar accents when speaking English, which makes them more confusable to speaker verification systems compared to learners from a different L1 group. Particularly, impostors of Thai L1 constitute a considerable portion of FA errors for each L1, as A1 and A2 speakers dominate Thai L1 in the BULATS test set, which is different from other L1s where B1 and B2 speakers dominate. 

\begin{table}[ht!]
  \caption{L1 breakdown of the percentage of impostor trials with FA errors at the operating threshold of EER for the extractor fine-tuning system on a subset of the BULATS test set.}
  \label{tab:L1_breakdown}
  \centering
  \vspace{1.5ex}
  \begin{tabular}{ccccccc}
    \hline
	L1 & \multicolumn{6}{c}{L1 of Impostor Spkr.} \\
	\cline{2-7}
	 Ref. Spkr.  & Ara. & Pol. & Spa. & Tam. & Tha. & Vie. \\
	 \hline
	 Ara. & \textbf{74.9} & 0.0 & 0.3 & 0.6 & 14.7 & 9.5 \\
	 Pol. & 0.0 & \textbf{76.9} & 1.3 & 0.3 & 21.6 & 0.0 \\
	 Spa. & 2.1 & 16.5 & \textbf{44.7} & 0.0 & 28.2 & 8.5 \\
	 Tam. & 0.0 & 1.7 & 0.3 & \textbf{62.4} & 33.9 & 1.7 \\
	 Tha. & 0.5 & 2.4 & 0.4 & 1.0 & \textbf{92.9} & 2.8 \\
	 Vie. & 1.2 & 0.1 & 1.3 & 0.6 & 12.7 & \textbf{84.0} \\
    \hline
  \end{tabular}
\end{table}

\subsection{Overall system performance}

Based on the analysis in the previous section, the impact of speaker attributes beyond gender, the grade and L1, were used as additional restrictions on the imposter set selection. The following forms of impostor selection were examined:
\begin{itemize}
    \item \textbf{gender}, impostors from the same gender group as the reference speaker, as in Section~\ref{sub:gender_dependent};
    \vspace{-.5ex}
    \item \textbf{grade}, impostors from the same grade group as the reference speaker;
    \item \textbf{$>$grade}, impostors from higher grade groups than the reference speaker if the grade of the reference speaker is lower than C, otherwise from C; this case is of practical interest for impersonation in spoken language tests;
    \item \textbf{L1}, impostors from the same L1 group as the reference speaker;
\end{itemize}

The number of total verification trials decreases with further restriction on impostors, which is shown in Table~\ref{tab:num_trials}. 
Table~\ref{tab:different_criteria} shows the impact on EER of restricting the possible set of impostors according to gender, L1 or grade on both BULATS and Linguaskill test sets. Due to the lack of data for each L1 or grade, X1 and X2 systems that are adapted or fine-tuned on all of the BULATS training set are used for verification. As expected, restricting possible impostors according to speaker attributes yielded higher EERs as the percentage of impostors ``close" to the reference speaker increased. Take \textbf{gender} as the starting point, which is the configuration used in previous experiments in Section~\ref{sub:gender_dependent}. Further restricting the set of impostors to \textbf{L1} again increased EERs agreeing with the results shown in Table~\ref{tab:L1_breakdown}, similarly to \textbf{grade}. An interesting result in terms of handling impersonation is that, if the set of impostors is further restricted to \textbf{$>$grade}, EERs decrease compared to simply restricted to \textbf{gender}. The highest EER for both systems was achieved by restricted to \textbf{gender}+\textbf{L1}+\textbf{grade}, which indicates that all these are important speaker attributes of non-native data.
The \textbf{gender}+\textbf{L1}+\textbf{$>$grade} case is more related to practical scenarios of impersonation, since it is more likely that a candidate chooses a substitute from the same gender and L1 group but speak the target language better to impersonate him/herself in order to obtain a higher grade in a spoken language test. 
\vspace{-2ex}

\begin{table}[h]
\caption{Number of verification trials (in millions) with different restrictions on impostors for both BULATS and Linguaskill test sets.} 
\label{tab:num_trials}
\centering{}%
\vspace{1.5ex}
\begin{tabular}{lcc}
\hline
 Restrictions & BULATS & Linguaskill \\
    \hline
    gender & 104.8 & 9.7 \\
	~~~~~~+ grade & 31.6 & 2.7 \\
	~~~~~~+ $>$grade & 36.9 & 3.6 \\
	~~~~~~+ L1 & 44.3 & 2.2 \\
	~~~~~~~~~~~~+ grade & 14.1 & 0.7 \\
	~~~~~~~~~~~~+ $>$grade & 16.7 & 0.8 \\
\hline
\end{tabular}
\end{table}

\vspace{-2ex}

\begin{table}[h]
\caption{\% EER performance of two adapted systems with different restrictions on impostors on both BULATS and Linguaskill test sets.} 
\label{tab:different_criteria}
\centering{}%
\vspace{1.5ex}
\begin{tabular}{lcccc}
\hline
 \multirow{2}{*}{Restrictions} & \multicolumn{2}{c}{BULATS} & \multicolumn{2}{c}{Linguaskill}\\
 \cline{2-5}
 & X1 & X2 & X1 & X2 \\
    \hline
    gender & 0.55 & 0.49 & 0.62 & 0.55 \\
	~~~~~~+ grade & 0.60 & 0.64 & 0.66 & 0.64 \\
	~~~~~~+ $>$grade & 0.45 & 0.49 & 0.55 & 0.49 \\
	~~~~~~+ L1 & 0.65 & 0.71 & 0.84 & 0.98 \\
	~~~~~~~~~~~~+ grade & 0.73 & 0.79 & 0.92 & 1.17 \\
	~~~~~~~~~~~~+ $>$grade & 0.62 & 0.68 & 0.79 & 0.87 \\
\hline
\end{tabular}
\end{table}

\begin{figure}[ht!]
  \centering
  \includegraphics[scale=0.5]{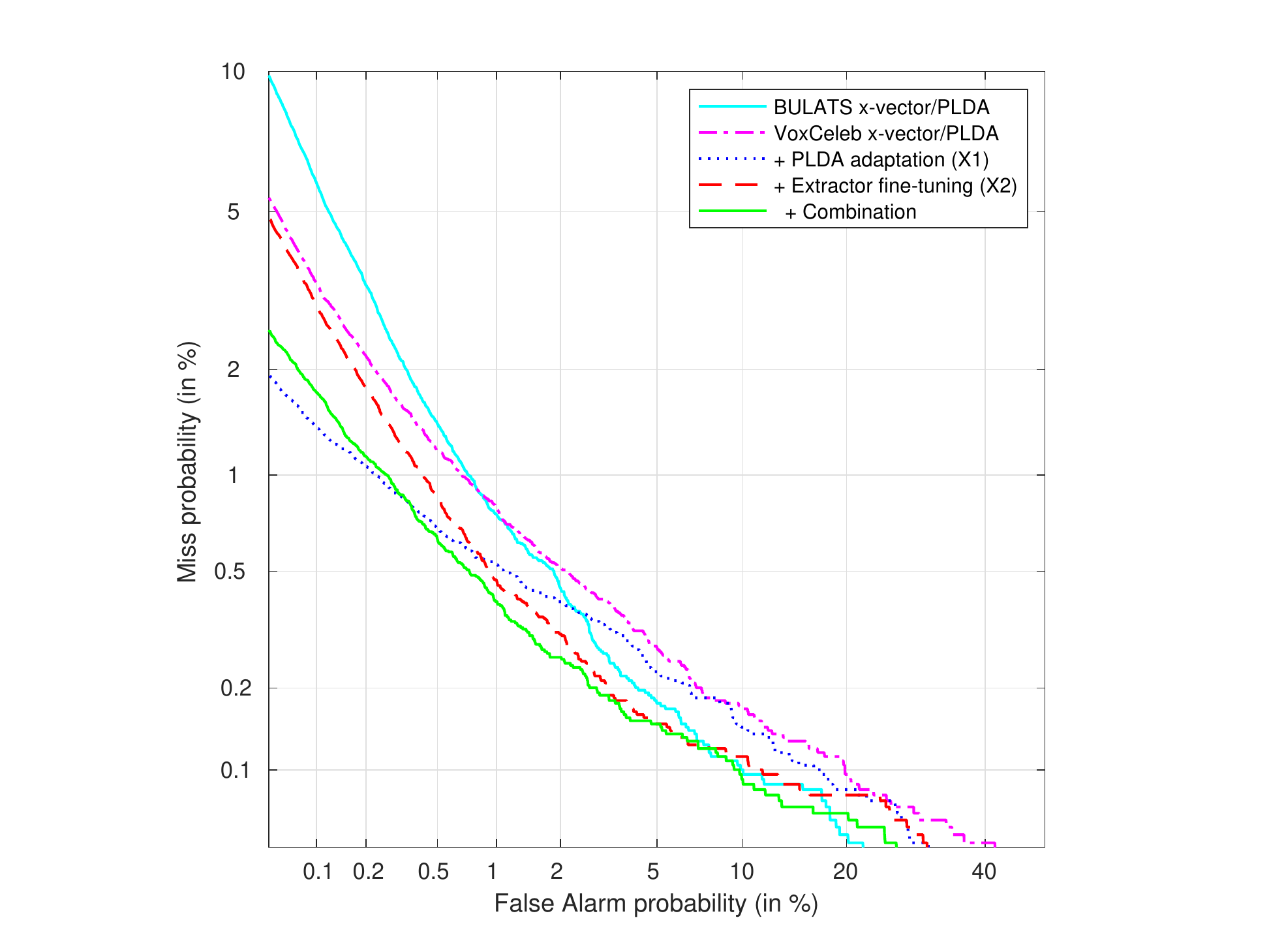}
  \caption{DET curves of various systems on the BULATS test set with impostor trials selected from the group of the same gender, same L1 and higher grade as/than the reference speaker. }
  \label{fig:det_curves}
  \vspace{-0.2cm}
\end{figure}

For the impersonation scenario where the impostor trials are restricted to \textbf{gender}+\textbf{L1}+\textbf{$>$grade}, the DET curves for all systems including the unadapted VoxCeleb and BULATS trained systems are shown in Figure~\ref{fig:det_curves} for the BULATS test set. This allows the overall distribution of FA and MS errors for the aforementioned systems to be evaluated. 
It can be seen that, compared with the fine-tuned X2 system, the PLDA-adapted X1 system had a lower MS probability when the FA probability was low and had a higher MS probability when the FA probability was high. This implies that the X1 system tends to accept imposters as reference speakers while the X2 system tends to reject reference speakers as impostors. For malpractice candidate impersonation in spoken language tests, the X2 system may have a high cost as it may incorrectly identify malpractice in valid candidates. This would require manual checks to confirm this classification. In contrast, the X1 system may result in a lower level of security because it has a higher chance of misidentifying the candidate who is impersonating another. Based on these complementary trends, a score-level linear combination of the two systems was performed with weights of 0.7 and 0.3 for X1 and X2 systems, respectively. The combination system gave consistently better performance for a wide range of FA and MS probabilities than the aforementioned systems with an EER of 0.58\% on the BULATS test set, as demonstrated in Figure~\ref{fig:det_curves}. The same trend was also observed at these weightings on the Linguaskill test set with an EER of 0.72\% for the combination system, approximately 8\% relative reduction in EER from the X1 system. Thus, the combination of the two adapted systems making use of both large-scale VoxCeleb data and in-domain BULATS data, can serve as a sensible configuration for impersonation detection in spoken language tests. 

\vspace{-2ex}
\vspace{-2ex}

\section{Conclusions}
This paper has investigated malpractice in the form of candidate impersonation for spoken language assessment. This task has close relationships to standard speaker verification, but applied to the domain of non-native speech.  Advanced  neural network based speaker verification systems were built on both limited non-native spoken English data from the BULATS test, and a large standard corpus VoxCeleb. For the configuration used all systems yielded relatively low EERs of less than 1\%. Though built with only limited data the systems trained on just BULATS systems outperformed the ``out-of-the-box" VoxCeleb based system. However by adapting both the PLDA model and the deep speaker representation, the VoxCeleb-based systems could yield lower EERs. The attributes of the ``impostors" was then analysed in terms of both the impostor's grade and L1. As expected, L1 was the most important attribute
of the impostor selected, though the grade did also influence performance. With the most likely scenario of impersonation by restricting impostors to be from the same gender, same L1, and higher grade group, the combination of the two adapted systems gave consistently better performance for a wide range of FA and MS probabilities, making it a sensible configuration for impersonation detection. 


\vfill
\pagebreak



\bibliographystyle{IEEEbib}
\bibliography{strings,refs}

\begin{thebibliography}{10}

\bibitem{Zechner2009}
K.~Zechner, D.~Higgins, X.~Xi, and D.~M. Williamson,
\newblock ``Automatic scoring of non-native spontaneous speech in tests of
  spoken {E}nglish,''
\newblock {\em Speech Communication}, vol. 51, no. 10, pp. 883--895, 2009.

\bibitem{Witt2000}
S.~M. Witt and S.~J. Young,
\newblock ``Phone-level pronunciation scoring and assessment for interactive
  language learning,''
\newblock {\em Speech Communication}, vol. 30, no. 2, pp. 95--108, 2000.

\bibitem{Metallinou2014}
A.~Metallinou and J.~Cheng,
\newblock ``{Using deep neural networks to improve proficiency assessment for
  children {E}nglish language learners},''
\newblock in {\em Proc. Interspeech}, 2014, pp. 1468--1472.

\bibitem{Wang2018a}
Y.~Wang, M.~J.~F. Gales, K.~M. Knill, K.~Kyriakipoulos, A.~Malinin, R.~C. van
  Dalen, and M.~Rashid,
\newblock ``Towards automatic assessment of spontaneous spoken {E}nglish,''
\newblock {\em Speech Communication}, vol. 104, pp. 47 -- 56, 2018.

\bibitem{Dehak2011}
N.~Dehak, P.~J. Kenny, R.~Dehak, P.~Dumouchel, and P.~Ouellet,
\newblock ``Front-end factor analysis for speaker verification,''
\newblock {\em IEEE Transactions on Audio, Speech, and Language Processing},
  vol. 19, no. 4, pp. 788--798, 2011.

\bibitem{Prince2007}
S.~J.~D. Prince and J.~H. Elder,
\newblock ``Probabilistic linear discriminant analysis for inferences about
  identity,''
\newblock in {\em IEEE International Conference on Computer Vision}, 2007, pp.
  1--8.

\bibitem{Kenny2010}
P.~Kenny,
\newblock ``Bayesian speaker verification with heavy-tailed priors.,''
\newblock in {\em Proc. Odyssey: Speaker and Language Recognition Workshop},
  2010, vol.~14.

\bibitem{Garcia2011}
D.~Garcia-Romero and C.~Y. Espy-Wilson,
\newblock ``Analysis of i-vector length normalization in speaker recognition
  systems,''
\newblock in {\em Proc. Interspeech}, 2011, pp. 249--252.

\bibitem{Garcia2012}
D.~Garcia-Romero, X.~Zhou, and C.~Y. Espy-Wilson,
\newblock ``Multicondition training of gaussian {PLDA} models in i-vector space
  for noise and reverberation robust speaker recognition,''
\newblock in {\em Proc. ICASSP}, 2012, pp. 4257--4260.

\bibitem{Variani2014}
E.~Variani, X.~Lei, E.~McDermott, I.~L. Moreno, and J.~Gonzalez-Dominguez,
\newblock ``Deep neural networks for small footprint text-dependent speaker
  verification,''
\newblock in {\em Proc. ICASSP}, 2014, pp. 4052--4056.

\bibitem{Snyder2016}
D.~Snyder, P.~Ghahremani, D.~Povey, D.~Garcia-Romero, Y.~Carmiel, and
  S.~Khudanpur,
\newblock ``Deep neural network-based speaker embeddings for end-to-end speaker
  verification,''
\newblock in {\em Proc. SLT Workshop}, 2016, pp. 165--170.

\bibitem{Snyder2017}
D.~Snyder, D.~Garcia-Romero, D.~Povey, and S.~Khudanpur,
\newblock ``Deep neural network embeddings for text-independent speaker
  verification.,''
\newblock in {\em Proc. Interspeech}, 2017, pp. 999--1003.

\bibitem{Snyder2018}
D.~Snyder, D.~Garcia-Romero, G.~Sell, D.~Povey, and S.~Khudanpur,
\newblock ``X-vectors: Robust {DNN} embeddings for speaker recognition,''
\newblock in {\em Proc. ICASSP}, 2018, pp. 5329--5333.

\bibitem{Shriberg2008}
E.~Shriberg, L.~Ferrer, S.~Kajarekar, N.~Scheffer, A.~Stolcke, and M.~Akbacak,
\newblock ``Detecting nonnative speech using speaker recognition approaches.,''
\newblock in {\em Odyssey: The Speaker and Language Recognition Workshop},
  2008.

\bibitem{Tan2010}
B.~Tan, Q.~Li, and R.~Foresta,
\newblock ``An automatic non-native speaker recognition system,''
\newblock in {\em IEEE International Conference on Technologies for Homeland
  Security (HST)}, 2010, pp. 77--83.

\bibitem{Omar2010}
M.~K. Omar and J.~Pelecanos,
\newblock ``A novel approach to detecting non-native speakers and their native
  language,''
\newblock in {\em Pro. ICASSP}, 2010, pp. 4398--4401.

\bibitem{Qian2016}
Y.~Qian et~al.,
\newblock ``{Self-Adaptive {DNN} for Improving Spoken Language Proficiency
  Assessment},''
\newblock in {\em Proc. Interspeech}, 2016.

\bibitem{Nagrani2017}
A.~Nagrani, J.~S. Chung, and A.~Zisserman,
\newblock ``Vox{C}eleb: A large-scale speaker identification dataset,''
\newblock in {\em Proc. Interspeech}, 2017, pp. 2616--2620.

\bibitem{Chung2018}
J.~S. Chung, A.~Nagrani, and A.~Zisserman,
\newblock ``Vox{C}eleb2: Deep speaker recognition,''
\newblock in {\em Proc. Interspeech}, 2018, pp. 1086--1090.

\bibitem{Povey2011}
D.~Povey, A.~Ghoshal, G.~Boulianne, L.~Burget, O.~Glembek, G.~Nagendra,
  M.~Hannemann, P.~Motl{\'i}{\v{c}}ek, Y.~Qian, P.~Schwarz, J.~Silovsk{\'y},
  G.~Stemmer, and K.~Vesel{\'y},
\newblock ``The {K}aldi speech recognition toolkit,''
\newblock in {\em Proc. ASRU Workshop}, 2011, pp. 4141--4144.

\bibitem{Heigold2016}
G.~Heigold, I.~Moreno, S.~Bengio, and N.~Shazeer,
\newblock ``End-to-end text-dependent speaker verification,''
\newblock in {\em Proc. ICASSP}, 2016, pp. 5115--5119.

\bibitem{Brummer2010}
N.~Br{\"u}mmer and E.~De~Villiers,
\newblock ``The speaker partitioning problem.,''
\newblock in {\em Proc. Odyssey: Speaker and Language Recognition Workshop},
  2010.

\bibitem{Ioffe2006}
S.~Ioffe,
\newblock ``Probabilistic linear discriminant analysis,''
\newblock in {\em European Conference on Computer Vision}. Springer, 2006, pp.
  531--542.

\bibitem{Garcia2014}
D.~Garcia-Romero, X.~Zhang, A.~McCree, and D.~Povey,
\newblock ``Improving speaker recognition performance in the domain adaptation
  challenge using deep neural networks,''
\newblock in {\em Proc. SLT Workshop}, 2014, pp. 378--383.

\bibitem{Garcia2014a}
D.~Garcia-Romero, A.~McCree, S.~Shum, N.~Br{\"u}mmer, and C.~Vaquero,
\newblock ``Unsupervised domain adaptation for i-vector speaker recognition,''
\newblock in {\em Proc. Odyssey: The Speaker and Language Recognition
  Workshop}, 2014.

\bibitem{Villalba2014}
J.~Villalba and E.~Lleida,
\newblock ``Unsupervised adaptation of {PLDA} by using variational {B}ayes
  methods,''
\newblock in {\em Proc. ICASSP}, 2014, pp. 744--748.

\bibitem{Garcia2014b}
D.~Garcia-Romero and A.~McCree,
\newblock ``Supervised domain adaptation for i-vector based speaker
  recognition,''
\newblock in {\em Proc. ICASSP}, 2014, pp. 4047--4051.

\bibitem{chambers-2011-bulats}
L.~Chambers and K.~Ingham,
\newblock ``The {BULATS} online speaking test,''
\newblock {\em Research Notes}, vol. 43, pp. 21--25, 2011.

\bibitem{CEFR2001}
{\relax Council of Europe},
\newblock {\em {C}ommon {E}uropean {F}ramework of {R}eference for {L}anguages:
  learning, teaching, assessment},
\newblock Cambridge University Press, 2001.

\bibitem{Young2015_htk}
S.~Young, G.~Evermann, M.~J.~F. Gales, T.~Hain, D.~Kershaw, X.~Liu, G.~Moore,
  J.~Odell, D.~Ollason, D.~Povey, A.~Ragni, V.~Valtchev, P.~C. Woodland, and
  C.~Zhang,
\newblock {\em The {HTK} book (for {HTK} version 3.5)},
\newblock University of Cambridge, 2015.

\end{thebibliography}

\end{document}